\documentclass[aps,prl,superscriptaddress,twocolumn]{revtex4}
\usepackage{amssymb}
\usepackage{graphicx}
\usepackage{amsmath}
\usepackage[colorlinks=true,linkcolor=blue,citecolor=blue,urlcolor=blue]{hyperref}

\begin{document}

\title{High Temperature Virial Expansion to Universal Quench Dynamics}

\author{Mingyuan Sun}
\email{mingyuansun@bupt.edu.cn}
\affiliation{State Key Lab of Information Photonics and Optical Communications, Beijing University of Posts and Telecommunications, Beijing 100876, China}
\affiliation{School of Science, Beijing University of Posts and Telecommunications, Beijing 100876, China}

\author{Peng Zhang}
\affiliation{Department of Physics, Renmin University of China, Beijing 100872,
China}
\affiliation{Beijing Key Laboratory of Opto-electronic Functional Materials \&
Micro-nano Devices, Renmin Univeristy of China, Beijing 100872, China}
\author{Hui Zhai}
\email{hzhai@tsinghua.edu.cn}
\affiliation{Institute for Advanced Study, Tsinghua University, Beijing 100084, China}
\date{\today}

\begin{abstract}

High temperature virial expansion is a powerful tool in equilibrium statistical mechanics. In this letter we generalize the high temperature virial expansion approach to treat far-from-equilibrium quench dynamics. As an application of our framework, we study the dynamics of a Bose gas quenched from non-interacting to unitarity, and we compare our theoretical results with unexplained experimental results by the Cambridge group [Eigen {\it et al.,} Nature {\bf 563}, 221 (2018)]. We show that, during the quench dynamics, the momentum distribution decreases for low-momentum part with $k<k^*$, and increases for high-momentum part with $k>k^*$, where $k^*$ is a characteristic momentum scale separating the low- and the high-momentum regimes. We determine the universal value of $k^*\lambda$ that agrees perfectly with the experiment, with $\lambda$ being the thermal de Broglie wave length. We also find a jump of the half-way relaxation time across $k^*\lambda$ and the non-monotonic behavior of energy distribution, both of which agree with the experiment. Finally, we address the issue whether the long-time steady state thermalizes or not, and we find that this state reaches a partial thermalization, namely, it thermalizes for low-energy part with $k\lambda\lesssim 1$ but not thermalizes for the very high momentum tail with $k\lambda\gg 1$. Our framework can also be applied to quench dynamics in other systems.       

\end{abstract}

\maketitle

Non-equilibrium quantum dynamics is nowadays a focused research topic in synthetic quantum systems like ultracold atomic gases. Starting from an equilibrium state, a sudden change of either the state or the Hamiltonian can bring the system out of equilibrium, and the subsequent evolution governed by the time-independent Hamiltonian is referred to as quench dynamics.  Quench dynamics is one of the most studied non-equilibrium dynamics, especially in ultracold atomic gases experiments \cite{quench_exp1,quench_exp2,quench_exp3,quench_exp4,quench_exp5}. This is because ultracold atomic gases are quite dilute such that the typical many-body time scales are about millisecond. Therefore, a typical operation at microsecond time scale is fast enough compared with the many-body time scales, which can be considered as a quench process. 

Regarding the goal of studying quench dynamics, one of the most important motivations is to extract universal features of a system from its quench dynamics. Here universal features refer to phenomena that are insensitive to the choice of the initial state and are connected to essential properties of the Hamiltonian governing the evolution. In the past few years, there are several established  examples of universal quench dynamics. For instance, the logarithmic growth versus the linear growth of the entanglement entropy after quench can distinguish many-body localization from thermalization \cite{MBL}. The linking number in the quench dynamics of a band insulator can reveal the topological invariant of the Hamiltonian \cite{linking_th,linking_exp1,linking_exp2}. Fractal structure in the time domain after quench can reveal the discrete scaling symmetry of the Hamiltonian \cite{Gao_Shi}.   

If a ultracold atomic gas is quenched from non-interacting to unitarity, where the $s$-wave scattering length diverges, one would expect to observe universal quench dynamics that are related to strongly interacting physics at unitarity. One important feature  of strongly interacting gases at unitarity is the emergence of scale invariance, which states that there is no extra length scale associated with interaction, therefore the inter-particle spacing and the thermal de Broglie wave length are only relevant length scales.  An experiment from the Cambridge group reported such universal quench dynamics of a Bose gas from non-interacting initial state to unitarity. Indeed, several universal phenomena have been observed in both the low-temperature condensed gas and the non-condensed thermal gas. This system has been studied by several theoretical papers before\cite{theory1,theory2,theory3,theory4,theory5,theory6,theory7,theory8,
theory9,theory10,theory11,theory12,Gao,Parish,Colussi}. The low temperature behavior has been calculated by using the Bogoliubov theory \cite{Gao,Parish}, and the universal exponential form observed in the momentum distribution has been explained \cite{Gao}. However, the thermal gas part of the data remains unexplained. The challenge, of course, lies in the absence of reliable theoretical tools to treat a strongly interacting system. 

In this letter we develop a framework to study quantum quench dynamics through high-temperature virial expansion. The advantage of virial expansion is that it uses fugacity as an expansion parameter and can be applied to strongly interacting regime \cite{virial}. Previously, virial expansion has been successfully used in studying unitary quantum gases \cite{virial_th1,virial_th2,virial_th3,virial_th4,virial_th5,virial_th6,
virial_th7,virial_th8,virial_th9,virial_th10,virial_th11,virial_th12,
virial_th13,virial_th14} and the results are compared well with experiments \cite{virial_exp1,virial_exp2,virial_exp3,virial_exp4,virial_exp5,
virial_exp6,virial_exp7,virial_exp8}, but these applications are limited to equilibrium properties. Our work generalizes this method to far-from-equilibrium situations. As an application of our theory, we consider the dynamics of momentum distribution when a Bose gas is quenched from non-interacting to unitarity, and we use our theoretical results to explain the thermal gas data of the Cambridge experiment. 

\textit{General Framework.} We consider a system initially at the equilibrium state of the free Hamiltonian $\hat{H}_0$ with temperature $T$, and starting from $t=0$, it evolves under an interacting Hamiltonian $\hat{H}$. When the system is evolved to time $t>0$, we measure the physical observable $\hat{W}$. This observable $\mathcal{W}(t)$ can be expressed as
\begin{equation}
    \mathcal{W}(t)=\frac{\text{Tr} [ e^{-\beta(\hat{H}_0-\mu \hat{N})}e^{it\hat{H}}\hat{W} e^{-it\hat{H}} ]}{\text{Tr}[e^{-\beta(\hat{H}_0-\mu \hat{N})} ]}
    \label{Wt}
\end{equation}
Here $\beta=1/(k_BT)$, and for convenience, we set $k_B=\hbar=1$ in this paper. $\mu$ and $N$ are the chemical potential and the total number of bosons, respectively.

At high temperature, the fugacity $z=e^{\beta\mu}\ll 1$, and we can expand both the numerator and the denominator in powers of $z$ as
\begin{equation}
  \mathcal{W}(t)=\frac{zX_1+z^2X_2+ \cdots}{1+zQ_1+z^2Q_2+ \cdots}.
  \label{Wt1}
\end{equation}
Here
\begin{align} 
X_n&=\text{Tr}_n[\Theta(t) e^{-\beta \hat{H}_0}e^{it\hat{H}}\hat{W} e^{-it\hat{H}}],\\
Q_n&=\text{Tr}_n[e^{-\beta \hat{H}_0}],
\end{align} 
where $n=0,1,2,...$ represents that the trace is taken over all eigen-states with totally $n$-number of particles. $\Theta(t)$ is the unit step function, with $\Theta(t)=1$ for $t\geq 0$ and $\Theta(t)=0$ otherwise. Note that one important difference between this virial expansion for dynamics and the virial expansion for equilibrium situation is that here $X_n$ depends on time $t$. Hence, the validity of this expansion requires that all $X_n$ do not exhibit divergent behavior as time evolves. 

$Q_n$ has been calculated for the virial expansion of equilibrium situation, and here the new task is to compute $X_n$. Let $\{|\psi_\alpha^{(n)}\rangle\}$ be a set of eigen-states for the non-interacting Hamiltonian with energy $E_\alpha^{(n)}$. By inserting the basis, it is straightforward to express $X_n$ as \begin{equation}
   X_n =\sum_{\alpha,\beta,\gamma}  e^{-\beta E_\alpha^{(n)}}G_{\beta\alpha}^{(n)*}(t) \langle \psi_\beta^{(n)}|   \hat{W}  |\psi_\gamma^{(n)} \rangle G_{\gamma\alpha}^{(n)}(t),     
    \label{Xn}
\end{equation}
where $G^{(n)}(t)$ is the retarded Green's function for the interacting Hamiltonian $\hat{H}$ of $n$-particle system, and  
\begin{align}
   G_{\gamma\alpha}^{(n)}(t) &=\langle \psi_\gamma^{(n)}| \Theta(t)e^{-itH} |\psi_\alpha^{(n)} \rangle  \nonumber \\
   &=\frac{i}{2\pi }\int_{-\infty}^\infty d\omega e^{-i\omega t} G_{\gamma\alpha}^{(n)}(\omega+i0^+).
     \label{Gt}
\end{align}
Therefore, through the virial expansion, the quench dynamics problem is translated into the properties of the retarded Green's function of the Hamiltonian after the quench. In this way, the evolution of the observable during quench dynamics can be used to infer universal properties of the quench Hamiltonian.

\begin{figure}[t] 
    \centering
    \includegraphics[width=0.45\textwidth]{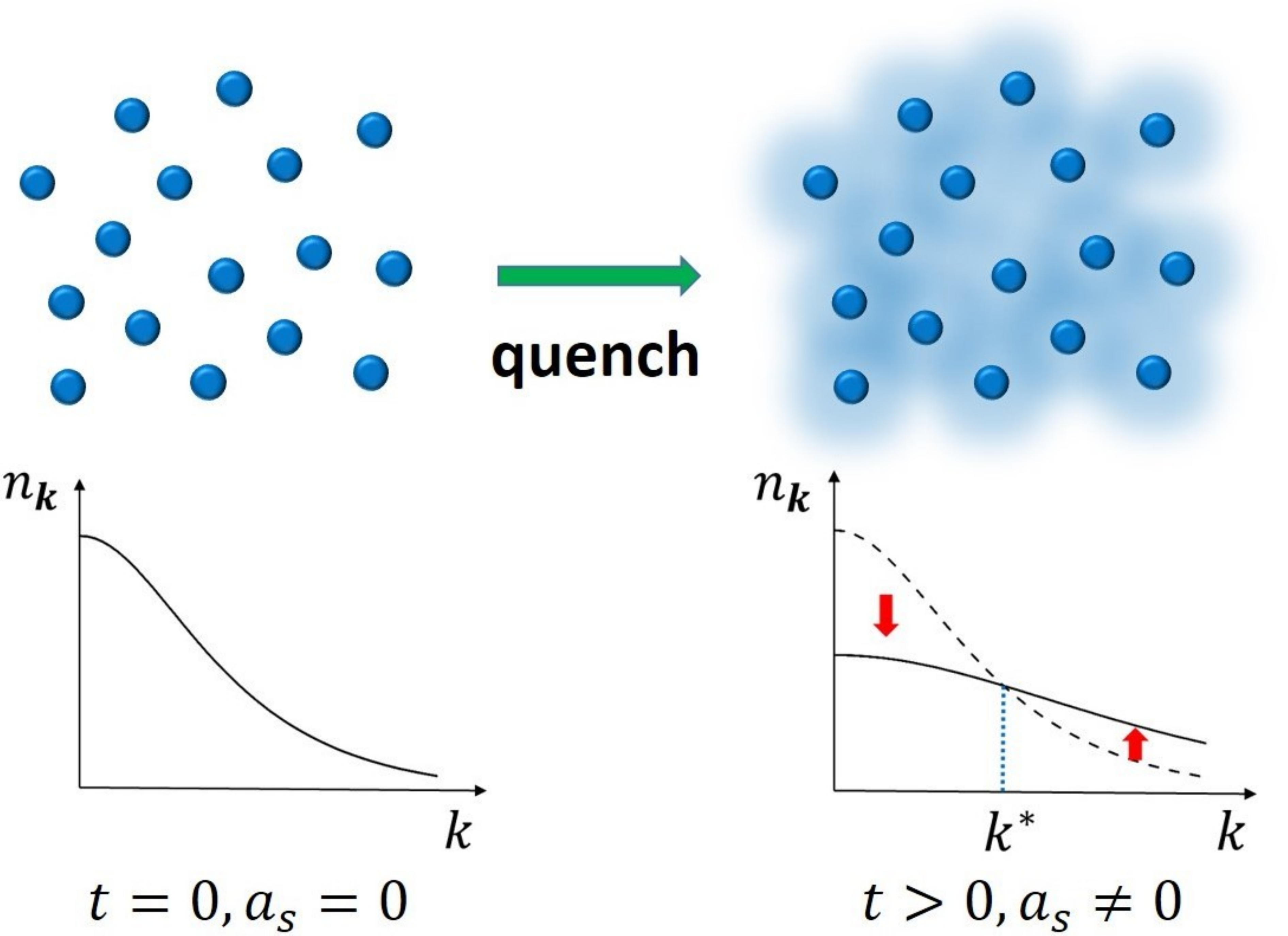}
    \caption{Cartoon picture of the quench dynamics of the Bose gas. The system is initially prepared at equilibrium of a non-interacting case with $a_\text{s}=0$, and at $t=0$, it is suddenly changed to an interacting case with $a_\text{s}\neq 0$, which remains at this situation afterward. In the subsequent evolution at $t>0$, the momentum distribution will become more broad. It schematically shows that, as time evolves, $n_{{\bf k}}$ decreases for small momentum $|{\bf k}|<k^*$ and $n_{{\bf k}}$ increases for large momentum $|{\bf k}|>k^*$. A characteristic crossover momentum is labelled by $k^*$.    }
     \label{fig1}
\end{figure}

As an application of our framework, below we consider a quench dynamics of three-dimensional uniform Bose gas from non-interacting to strongly interacting, as shown in Fig. \ref{fig1}. Here $\hat{H}_0=\sum_{i}{\bf p}_i^2/(2m)$ is the non-interacting Hamiltonian, with $m$ being the mass of bosons. At $t=0$, the Hamiltonian is suddenly changed to $\hat{H}=\hat{H}_0+\sum_{i<j}V(|{\bf r}_i-{\bf r}_j|)$, where $V(r)$ is the short-range interaction potential between bosons, described by the scattering length $a_\text{s}$. 
For the demonstration purpose, we focus on virial expansion up to $n=2$, and we will see below that reasonably good agreements between theory and experiment can already be obtained at this order. 

In this case, each two-body eigenstate is labelled by two quantum number $|\psi_{\alpha}^{(2)} \rangle=|{\bf P},{\bf q} \rangle$, with ${\bf P}$ and ${\bf q}$ being the total momentum and the relative momentum of two bosons, respectively. For energy of the free Hamiltonian, we have 
\begin{equation}
E_\alpha^{(2)}=\frac{{\bf P}^2}{4m}+\frac{{\bf q}^2}{m}, \label{free-energy}
\end{equation} and the two-body retarded Green's function can be expressed as \cite{Taylor}
\begin{equation}
   G_{\alpha\beta}^{(2)}(s) =\left[ \frac{\langle {\bf q_1}|{\bf q_2}\rangle }{s-\varepsilon_{\bf q_1}}+ \frac{T_2(s)}{(s-\varepsilon_{\bf q_1})(s-\varepsilon_{\bf q_2})}\right]\delta_{{\bf P}_1,{\bf P}_2}, \label{green}
\end{equation}
with $\alpha=\{{\bf P}_1,{\bf q_1}\}$, $\beta=\{{\bf P}_2,{\bf q_2}\}$, $s=\omega+i0^+$ and $\varepsilon_{\bf q}={\bf q}^2/m$. The two-body scattering T-matrix $T_2(s)$ is given by 
\begin{equation}
T_2(s)=\frac{4\pi/m}{a_\text{s}^{-1}-\sqrt{-ms}}. \label{T2}
\end{equation} 
For the interest of ultracold atom experiments, the time-of-flight measurement can directly measure $n_{{\bf k}}$. Hence, we consider $\hat{W}=\hat{n}_{{\bf k}}$, and
\begin{equation}
\langle \psi^{(2)}_\alpha|\hat{n}_{{\bf k}}|\psi^{(2)}_\beta\rangle=\left(\delta_{{\bf k},\frac{{\bf P_1}}{2}+{\bf q_1}}+\delta_{{\bf k},\frac{{\bf P_1}}{2}-{\bf q_1}}\right)\delta_{{\bf q_1}{\bf q_2}}\delta_{{\bf P_1}{\bf P_2}}.
\label{observable}
\end{equation}
With Eq. \ref{free-energy},\ref{green},\ref{T2},\ref{observable}, we complete all required terms in $X_2$ given by Eq. \ref{Xn}. Moreover, with the help of Eq. \ref{Wt1}, we can compute $n_{{\bf k}}(t)$ and we determine the change of momentum distribution as $\delta n_{{\bf k}}(t)=n_{{\bf k}}(t)-n_{{\bf k}}(0)$. 

\begin{figure}[t] 
    \centering
    \includegraphics[width=0.42\textwidth]{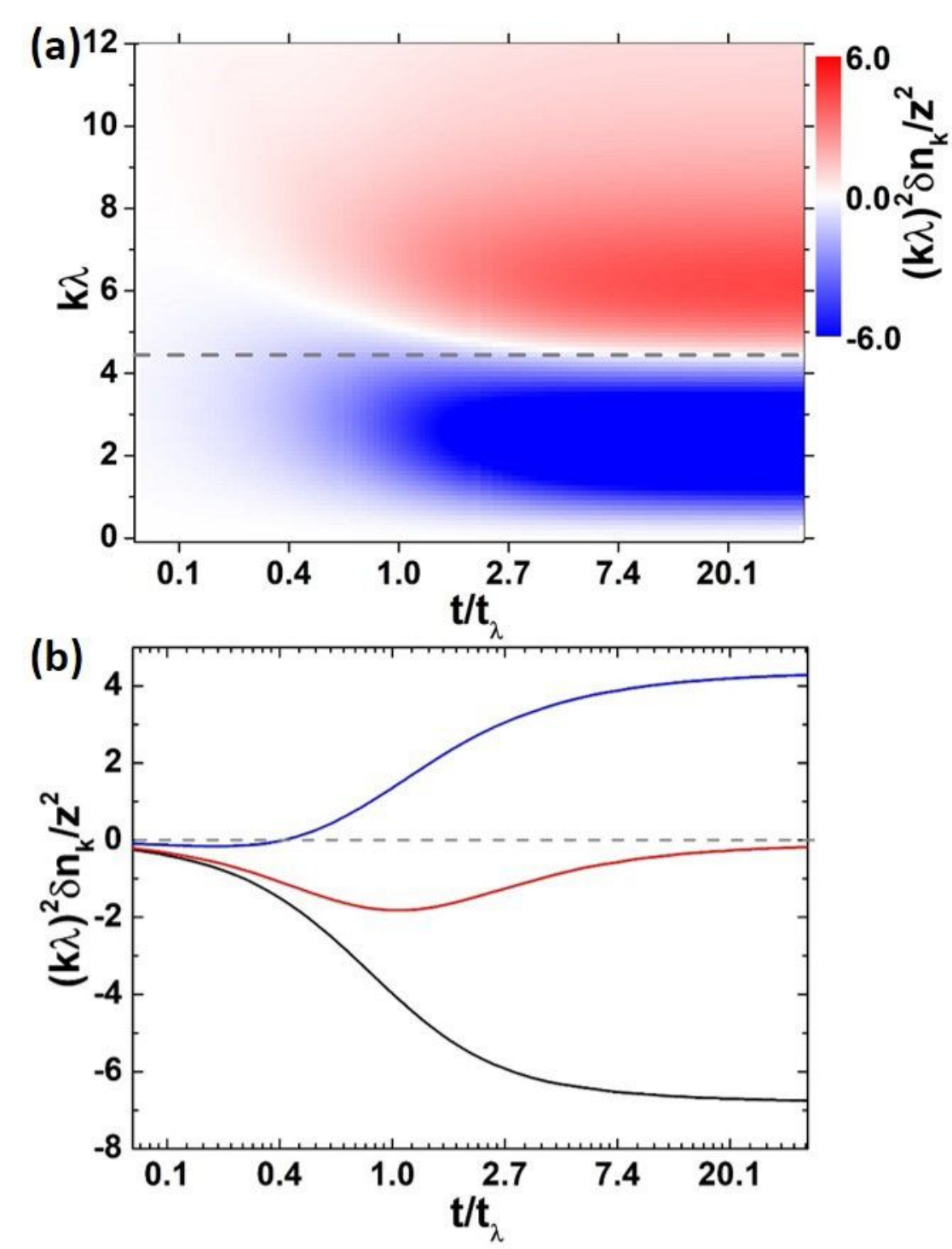}
    \caption{The evolution of the momentum distribution after the quench. (a) $k^2\delta n_{{\bf k}}$ (in unit of $z^2/\lambda^2$) as a function of $k\lambda$ and $t/t_\lambda$. (b) $k^2\delta n_{{\bf k}}$ (in unit of $z^2/\lambda^2$) as a function of $t/t_\lambda$ for $k\lambda=3.5$ ($k<k^*$, the black line),  $k\lambda=4.5$ ($k\sim k^*$, the red line) and $k\lambda=6$ ($k>k^*$, the blue line). }
     \label{fig2}
\end{figure}

\textit{Numerical Results and Experimental Comparison.} Here we consider the quench problem of a Bose gas from non-interacting case with $a_\text{s}=0$ to unitary regime with $a_\text{s}=\infty$, and we numerically solve the virial expansion up to $z^2$ order. In the virial expansion, our natural length unit is the thermal de Broglie wave length $\lambda$ and the energy unit is $T$. We also define a time unit $t_\lambda=1/T$. We will compare our calculation with the experimental observation from the Cambridge group. 

First of all, when the system is suddenly quenched to strongly interacting regime, atoms are scattered from low-momentum to high-momentum such that the increased interaction energy is converted into the kinetic energy. Hence, $n_{{\bf k}}$ decreases as a function of time $t$ for low-momentum, say, for $|{\bf k}|< k^*$, and $n_{{\bf k}}$ increases as a function of time $t$ for high-momentum, say, for $|{\bf k}|>k^*$. Here we use $k^*$ to denote a crossover momentum scale between low- and high-momentum. This physical picture is schematically shown in Fig. \ref{fig1}, and we can also see this clearly in Fig. \ref{fig2}(a), in which we plot $|{\bf k}|^2\delta n_{{\bf k}}$ as a function of time $t/t_\lambda$ and momentum $k\lambda$. Note that $|{\bf k}|^2\delta n_{{\bf k}}$ counts the change of occupation in a momentum shell with radius $|{\bf k}|$. There are two issues regarding $k^*$. One is what the value of the crossover momentum scale $k^*$ is, and the other is what the behavior of $n_{{\bf k}}$ for $|{\bf k}|\sim k^*$ is. These two questions are answered by Fig. \ref{fig2}(b). It first shows that $n_{{\bf k}}$ either monotonically decreases for $|{\bf k}|<k^*$ or monotonically increases for $|{\bf k}|>k^*$. For $|{\bf k}|\sim k^*$, our calculation shows that $n_{{\bf k}}$ first decreases slightly and then increases back to its initial value. This phenomenon agrees with what has been seen in experimental data. Here we use the condition $\delta n_{{\bf k}}(t\rightarrow \infty)=0$ to determine $k^*$ and we find $k^*\lambda=4.5$. This agrees remarkably well with the experimental result, where $k^*\lambda=4.4$.  

\begin{figure}[t] 
    \centering
    \includegraphics[width=0.42\textwidth]{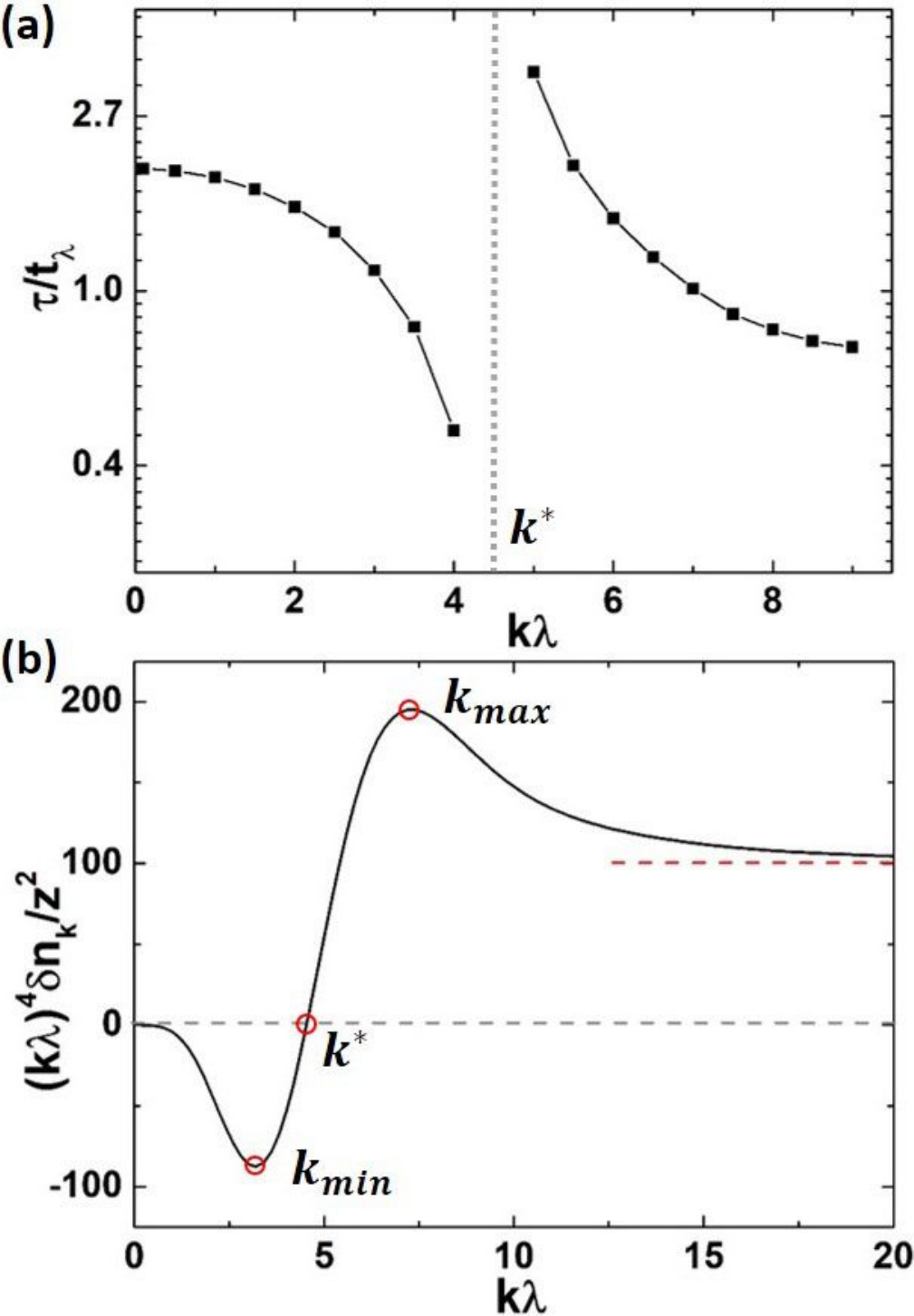}
    \caption{(a) The half-way time $\tau$ (see text for the definition) for different momenta $k\lambda$. There is no data around $k\lambda\sim 4.5$, where the half-way time is not well-defined due to the non-monotonic behavior of $\delta n_{{\bf k}}(t)$. (b) The variation of kinetic energy density $k^4\delta n_{{\bf k}}$ at final state with $t\rightarrow \infty$. This value is also plotted as a function of $k\lambda$. $k_\text{min}\lambda$ and $k_\text{max}\lambda$ are labelled where the minimum and the maximum of this function are taken. $k^*\lambda$ locates at the zero crossing. The red dashed line denotes the Contact $\mathcal{C}(t\rightarrow\infty)=32\pi z^2/\lambda^4$, which can be derived analytically.}
     \label{fig3}
\end{figure}

Secondly, in order to calibrate how fast the momentum distribution relaxes to its long-time equilibrium value, we introduce the half-way time $\tau_{{\bf k}}$ for each ${\bf k}$, which is defined as $\delta n_{{\bf k}}(\tau_{{\bf k}})=\frac{1}{2}\delta n_{{\bf k}}(t\rightarrow \infty)$. A larger $\tau_{{\bf k}}$ means a longer time for the time evolution to reach saturation. Note that $\tau_{{\bf k}}$ is not well-defined for $|{\bf k}|\sim k^*$, where $\delta n_{{\bf k}}(t)$ is not a monotonic function. In Fig. \ref{fig3}(a) we plot $\tau_{{\bf k}}$ as a function of $k\lambda$, except for the vicinity of $k^*\lambda$. It shows that for both $k<k^*$ regime and $k>k^*$ regime, $\tau_{{\bf k}}$ decreases as $k$ increases. However, there is a jump of $\tau_{{\bf k}}$ across $k\sim k^*$. This discontinuity of $\tau_{{\bf k}}$ is another interesting feature of the crossover momentum $k^*$ in this model. This feature shows that the relaxation time for momentum slightly above $k^*$ is much longer than the relaxation time for momentum slightly below $k^*$. This feature also agrees very well with experimental observation from the Cambridge group, and it is not obvious prior to the calculation.   

Thirdly, we plot the saturation value of $|{\bf k}|^4\delta n_{{\bf k}}$ taken at $t\rightarrow \infty$, as a function of $k\lambda$. Since $|{\bf k}|^2\delta n_{{\bf k}}$ plotted in Fig. \ref{fig2} represents the change of occupation in a momentum shell with radius $|{\bf k}|$, $|{\bf k}|^4\delta n_{{\bf k}}$ plotted in Fig. \ref{fig3}(b) represents the change of total kinetic energy in the same momentum shell. This function displays following behavior. It first decreases to a minimum located at $k_\text{min}\lambda$, and then it increases to a maximum located at $k_\text{max}\lambda$. Finally it decreases and saturates at large momentum. This functional behavior also agrees very well with experimental data. In our calculation, we find $k_\text{min}\lambda=3.2$ and $k_\text{max}\lambda=7.4$, which agree remarkably well with the experimental value $k_\text{min}\lambda\approx 3.2$ and $k_\text{max}\lambda\approx 7.5$. This function crosses zero naturally at $k^*\lambda$ because $k^*$ is defined as $\delta n_{|{\bf k}|=k^*}(t \rightarrow\infty)=0$. 

\begin{figure}[t] 
    \centering
    \includegraphics[width=0.42\textwidth]{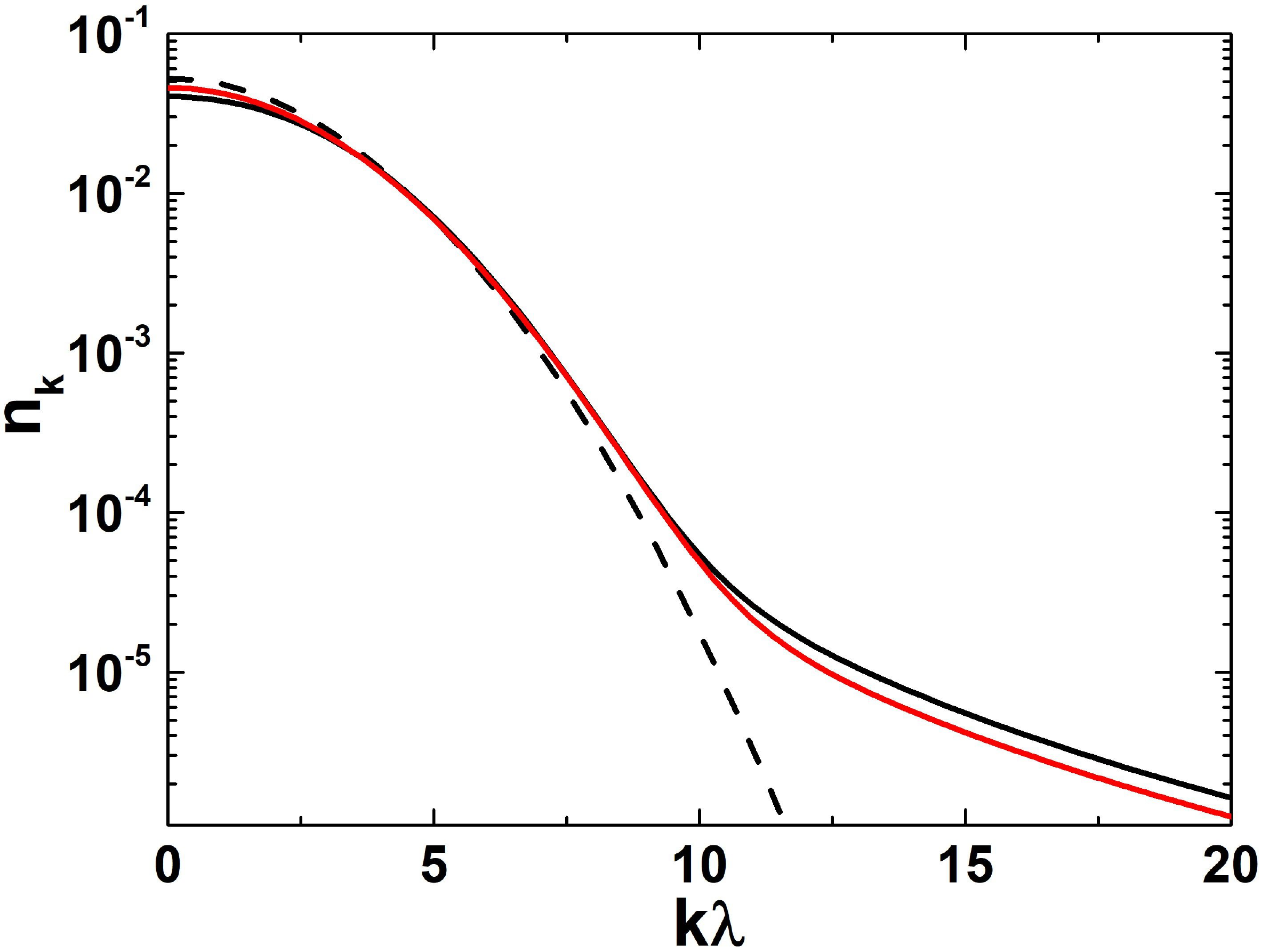}
    \caption{Comparison between $n_{{\bf k}}(t\rightarrow\infty)$ (the black solid line) with $z=0.05$ and an equilibrium momentum distribution $n^\text{eff}_{{\bf k}}$ (the red solid line). The dashed line denotes $n_{{\bf k}}(t=0)$.   }
     \label{fig4}
\end{figure}

Here we should make two comments regarding comparison with the experiment. First of all, in the experimental plots similar as our Fig. \ref{fig3}, they have used $E_\text{n}=\hbar^2 k^2_\text{n}/(2m)$ with $k_\text{n}=(6\pi^2 n)^{1/3}$ in the energy unit and $t_\text{n}=1/E_\text{n}$ in the time unit, in order to obtain universal scalings. They are different from our energy and time units. This is because, although the experiment is carried out in a non-condensed gas, the temperature is still close to the condensation temperature and higher order contributions in the virial expansion should be considered systematically. In our calculation, when the fugacity $z$ and temperature $T$ are fixed, $t_\lambda/t_\text{n}$ is a constant. Therefore, these two different choices of units only differ by a scale factor. Hence, the features we have discussed above, including the values of $k^*\lambda$, $k_\text{min}\lambda$ and $k_\text{max}\lambda$, the jump of $\tau$ and the non-monotonic behavior of energy distribution, are not affected by the change of units. It is quite remarkable to see that these main features can already be captured fairly well in the second order expansion. Secondly, our calculation also ignores the three-body inelastic contributions, which can cause loss and heating in real experiment. Fortunately, it turns out the time scale for the loss dynamics is larger than the time scale for relaxation by two-body collision \cite{rate1,rate2,quench_exp1,quench_exp2}. Therefore, experimentally, one can observe a steady state in a time window before the system is finally heated up by inelastic collisions. This steady state is what we compare with. 

\textit{Thermalization.} As we have shown that several key features of the quench dynamics obtained by the virial expansion agree well with the steady state observed in the experiment, we shall further ask a question that whether this steady state thermalizes. Here we consider the momentum distribution $n_{{\bf k}}(t\rightarrow\infty)$ after sufficiently long time evolution, as shown by the black solid line in Fig. \ref{fig4}. The question is whether we can find out an effective temperature, corresponding to $\lambda_\text{eff}$, and an effective chemical potential, corresponding to fugacity $z_\text{eff}$, with which the thermal equilibrium momentum distribution of a unitary Bose gas can reproduce $n_{{\bf k}}(t\rightarrow\infty)$.   

To determine $z_\text{eff}$ and $\lambda_\text{eff}$, we employ the energy conservation and the number conservation. Right after quench, the energy density of the system is determined by
\begin{equation}
\mathcal{E}=\frac{1}{V}\text{Tr}(e^{-\beta\hat{H}_0}\hat{H}). \label{E_initial}
\end{equation}
It is straightforward to calculate Eq. \ref{E_initial} using the second-order virial expansion, and we obtain
\begin{equation}
\mathcal{E}=\frac{3\pi}{m\lambda^5}z+\frac{3\sqrt{2}\pi}{8m\lambda^5}z^2. \label{E_initial_explicity}
\end{equation} 
Note that here $z$ and $\lambda$ are fugacity and thermal de Broglie wave length of initial non-interacting state, respectively.
And for a unitary Bose gas at equilibrium, we have energy density given by 
\begin{equation}
\mathcal{E}=\frac{3\pi}{m\lambda^5_\text{eff}}z_\text{eff}+\frac{27\sqrt{2}\pi}{8m\lambda^5_\text{eff}}z_\text{eff}^2. \label{E_equalibrium}
\end{equation} 
Hence we obtain the first equation by equalling Eq. \ref{E_initial_explicity} with Eq. \ref{E_equalibrium}.
Then, we note that the atom number for initial non-interacting gas is given by
\begin{equation}
n=\frac{1}{\lambda^3}\left(z+\frac{z^2}{2\sqrt{2}}\right), \label{n_0}
\end{equation}
and the equilibrium density of a unitary Bose gas is given by 
\begin{equation}
n=\frac{1}{\lambda^3_\text{eff}}\left(z_\text{eff}+\frac{9z^2_\text{eff}}{2\sqrt{2}}\right). \label{n_1}
\end{equation} 
By equalling Eq. \ref{n_0} and Eq. \ref{n_1}, we obtain the second equation. With these two equations, we can determine $z_\text{eff}$ and $\lambda_\text{eff}$ for a given initial $z$ and $\lambda$. 

Furthermore, we can obtain an equilibrium momentum distribution $n^\text{eff}_{{\bf k}}$ using the second virial expansion with $z_\text{eff}$ and $\lambda_\text{eff}$, as shown by the red solid line in Fig. \ref{fig4}. In Fig. \ref{fig4} we compare $n^\text{eff}_{{\bf k}}$ with $n_{{\bf k}}(t\rightarrow\infty)$. We can see that these two distributions agree very well as long as $k\lambda<10$. Visible relative  deviation can be seen for $k\lambda>10$, although $n_{{\bf k}}(t\rightarrow\infty)$ itself is already very small in this high energy tail. This indicates that the system thermalizes except for the high-energy tail. It can be understood in term of two-body collision section, which behaves as $\sim 1/k^2$ at unitarity. Thus, when $k^2/(2m)\gg k_\text{b}T$, the two-body collision section is too small to ensure thermalization.   

\textit{Conclusion.} In summary, we develop a virial expansion framework to study far-from-equilibrium quench dynamics. In the second-order virial expansion, we show that the results can already explain a number of experimental observations. By systematically including high-order contributions, we can also investigate the manifestation of the Efimov effect in the quench dynamics. Our framework can also be applied to study dynamics in other strongly interacting systems, such as the Bose and Fermi Hubbard model in optical lattices. 

\textit{Acknowledgment.} We thank Wei Zheng, Zheyu Shi, Ran Qi, Xin Chen, Chao Gao, Zhigang Wu for inspiring discussion.
The project was supported by Fund of State Key Laboratory of IPOC (BUPT) (No. 600119525, 505019124), NSFC Grant No. 11734010,  Beijing Outstanding Young Scholar Program and MOST under Grant No. 2016YFA0301600.


\begin{thebibliography}{99}



\bibitem{quench_exp1}
P. Makotyn, C. E. Klauss, D. L. Goldberger, E. A. Cornell, and D. S. Jin, Nat. Phys. {\bf 10}, 116 (2014).

\bibitem{quench_exp2}
C. Eigen, {\it et al}, Phys. Rev. Lett. {\bf 119}, 250404 (2017).

\bibitem{quench_exp3}
C. Eigen, J. A. Glidden, R. Lopes, E. A. Cornell, R. P. Smith, and Z. Hadzibabic, Nature {\bf 563}, 221 (2018).

\bibitem{quench_exp4}
M. Prufer, P. Kunkel, H. Strobel, S. Lannig, D. Linnemann, C. M. Schmied, J. Berges, T. Gasenzer, and M. K. Oberthaler, Nature {\bf 563}, 217 (2018).

\bibitem{quench_exp5}
S. Erne, R. Bucker, T. Gasenzer, J. Berges, and J. Schmiedmayer, Nature {\bf 563}, 225 (2018).


\bibitem{MBL}
D. A. Abanin, E. Altman, I. Bloch, and M. Serbyn, Rev. Mod. Phys. {\bf 91}, 021001 (2019).


\bibitem{linking_th}
C. Wang, P. F. Zhang, X. Chen, J. L. Yu, and H. Zhai, Phys. Rev. Lett. {\bf 118}, 185701 (2017).

\bibitem{linking_exp1}
M. Tarnowski, F. N. Unal, N. Fläschner, B. S. Rem, A. Eckardt, K. Sengstock, and C. Weitenberg, Nat. Commun. {\bf 10}, 1728 (2019).

\bibitem{linking_exp2}
W. Sun, {\it et al}, Phys. Rev. Lett. {\bf 121}, 250403 (2018).



\bibitem{Gao_Shi}
C. Gao, H. Zhai, and Z. Y. Shi, Phys. Rev. Lett. {\bf 122}, 230402 (2019).

\bibitem{theory1}
X. Yin and L. Radzihovsky, Phys. Rev. A {\bf 88}, 063611 (2013).
\bibitem{theory2}
A. G. Sykes, J. P. Corson, J. P. D'Incao, A. P. Koller, C. H. Greene, A. M. Rey, K. R. Hazzard, and J. L. Bohn, Phys. Rev. A {\bf 89}, 021601 (2014).
\bibitem{theory3}
A. Ran{\c{c}}on and K. Levin, Phys. Rev. A {\bf 90}, 021602 (2014).
\bibitem{theory4}
B. Kain and H. Y. Ling, Phys. Rev. A {\bf 90}, 063626 (2014).
\bibitem{theory5}
J. P. Corson and J. L. Bohn, Phys. Rev. A {\bf 91}, 013616 (2015).
\bibitem{theory6}
F. Ancilotto, M. Rossi, L. Salasnich, and F. Toigo, Few-Body Syst. {\bf 56}, 801 (2015).
\bibitem{theory7}
X. Yin and L. Radzihovsky, Phys. Rev. A {\bf 93}, 033653 (2016).
\bibitem{theory8}
V. E. Colussi, J. P. Corson, and J. P. D'Incao, Phys. Rev. Lett. {\bf 120}, 100401 (2018).
\bibitem{theory9}
V. E. Colussi, S. Musolino, and S. J. J. M. F. Kokkelmans, Phys. Rev. A {\bf 98}, 051601 (2018).
\bibitem{theory10}
M. Van Regemortel, H. Kurkjian, M. Wouters, and I. Carusotto, Phys. Rev. A {\bf 98}, 053612 (2018).
\bibitem{theory11}
J. P. D'Incao, J. Wang, and V. E. Colussi, Phys. Rev. Lett. {\bf 121}, 023401 (2018).
\bibitem{theory12}
S. Musolino, V. E. Colussi, and S. J. J. M. F. Kokkelmans, Phys. Rev. A {\bf 100}, 013612 (2019).

\bibitem{Gao}
C. Gao, M. Y. Sun, P. Zhang, H. Zhai, Phys. Rev. Lett. {\bf 124}, 040403 (2020).

\bibitem{Parish}
A. Mu\~noz de las Heras, M. M. Parish, F. M. Marchetti, Phys. Rev. A {\bf 99}, 023623 (2019).

\bibitem{Colussi}
V. E. Colussi, B. E. van Zwol, J. P. D'Incao, and S. J. J. M. F. Kokkelmans, Phys. Rev. A {\bf 99}, 043604 (2019).


\bibitem{virial}
K. Huang, {\it Statistical Mechanics}, (John Wiley \& Sons, New York, 1987).



\bibitem{virial_th1}
T. L. Ho and E. J. Mueller, Phys. Rev. Lett. {\bf 92}, 160404 (2004).

\bibitem{virial_th2}
X. Liu, H. Hu, and P. D. Drummond, Phys. Rev. Lett. {\bf 102},
160401 (2009).

\bibitem{virial_th3}D. B. Kaplan and S. Sun, Phys. Rev. Lett. {\bf 107}, 030601 (2011).

\bibitem{virial_th4}X. Leyronas, Phys. Rev. A {\bf 84}, 053633 (2011). 

\bibitem{virial_th5}
X. J. Liu, Phys. Rep. {\bf 524}, 37 (2013).

\bibitem{virial_th6} V. Ngampruetikorn, J. Levinsen, and M. M. Parish, Phys. Rev. Lett. {\bf 111}, 265301 (2013). 

\bibitem{virial_th7}M. Barth and J. Hofmann, Phys. Rev. A {\bf 89}, 013614 (2014). 

\bibitem{virial_th8}M. Sun and X. Leyronas, Phys. Rev. A {\bf 92}, 053611 (2015). 

\bibitem{virial_th9}M. Barth and J. Hofmann, Phys. Rev. A {\bf 92}, 062716 (2015). 

\bibitem{virial_th10}
M. Y. Sun, H. Zhai, and X. L. Cui, Phys. Rev. Lett. \textbf{119}, 013401 (2017).

\bibitem{virial_th11}
M. Y. Sun and X. L. Cui, Phys. Rev. A \textbf{96}, 022707 (2017).


\bibitem{virial_th12}
T. Enss, Phys. Rev. Lett. {\bf 123}, 205301 (2019).

\bibitem{virial_th13}
Y. Nishida, Ann. Phys. (Amsterdam) {\bf 410}, 167949 (2019).

\bibitem{virial_th14}
J. Hofmann, Phys. Rev. A {\bf 101}, 013620 (2020).




\bibitem{virial_exp1}
T. Bourdel, {\it et al}, Phys. Rev. Lett. {\bf 91}, 020402 (2003).


\bibitem{virial_exp2}
J. T. Stewart, J. P. Gaebler, D. S. Jin, Nature {\bf 454}, 744 (2008).

\bibitem{virial_exp3}
S. Nascimbéne, N. Navon, K. J. Jiang, F. Chevy, C. Salomon, Nature {\bf 463}, 1057 (2010).

\bibitem{virial_exp4}
E. D. Kuhnle, S. Hoinka, P. Dyke, H. Hu, P. Hannaford, and C. J. Vale, Phys. Rev. Lett. {\bf 106}, 170402 (2011).

\bibitem{virial_exp5}
M. Feld, B. Fröhlich, E. Vogt, M. Koschorreck, and M. Köhl, Nature (London) {\bf 480}, 75 (2011).

\bibitem{virial_exp6}
M. J. H. Ku, A. T. Sommer, L. W. Cheuk, M. W. Zwierlein, Science {\bf 335}, 563 (2012).

\bibitem{virial_exp7}
B. Mukherjee, P. B. Patel, Z. Yan, R. J. Fletcher, J. Struck, and M. W. Zwierlein, Phys. Rev. Lett. {\bf 122}, 203402 (2019).

\bibitem{virial_exp8}
C. Carcy, S. Hoinka, M. G. Lingham, P. Dyke, C. C. N. Kuhn, H. Hu, and C. J. Vale, Phys. Rev. Lett. {\bf 122}, 203401 (2019).


\bibitem{Taylor}
J. R. Taylor, {\it Scattering Theory} (Wiley, New York, 1972), Chapter 8.


\bibitem{rate1}
B. S. Rem,{\it et al}, Phys. Rev. Lett. {\bf110}, 163202 (2013).

\bibitem{rate2}
R. J. Fletcher, A. L. Gaunt, N. Navon, R. P. Smith, and Z. Hadzibabic, Phys. Rev. Lett. {\bf111}, 125303 (2013).




\end{thebibliography}
\end{document}